\documentclass[11pt]{article}

\let\counterwithin\relax

\usepackage{amsmath, amsfonts, amssymb, amsthm, bm, graphicx, mathtools, enumerate,multirow}
\usepackage{natbib}
\usepackage[CJKbookmarks=true,
bookmarksnumbered=true,
bookmarksopen=true,
colorlinks=true,
citecolor=blue,
linkcolor=blue,
anchorcolor=blue,
urlcolor=blue]{hyperref}
\usepackage[usenames]{color}
\usepackage[letterpaper, left=1.2truein, right=1.2truein, top = 1.2truein,
bottom = 1.2truein]{geometry}
\usepackage[ruled, lined, commentsnumbered]{algorithm2e}
\usepackage{prettyref,soul}

\usepackage{apptools} 
\usepackage{chngcntr} 
\AtAppendix{\counterwithin{lemma}{section}} 
\usepackage{tikz}
\usepackage[font=small,labelfont=bf]{caption}
\usepackage{enumitem}

\theoremstyle{definition}

\newrefformat{eq}{(\ref{#1})}
\newrefformat{chap}{Chapter~\ref{#1}}
\newrefformat{sec}{Section~\ref{#1}}
\newrefformat{algo}{Algorithm~\ref{#1}}
\newrefformat{fig}{Fig.~\ref{#1}}
\newrefformat{tab}{Table~\ref{#1}}
\newrefformat{rmk}{Remark~\ref{#1}}
\newrefformat{clm}{Claim~\ref{#1}}
\newrefformat{def}{Definition~\ref{#1}}
\newrefformat{cor}{Corollary~\ref{#1}}
\newrefformat{lmm}{Lemma~\ref{#1}}
\newrefformat{lemma}{Lemma~\ref{#1}}
\newrefformat{prop}{Proposition~\ref{#1}}
\newrefformat{app}{Appendix~\ref{#1}}
\newrefformat{ex}{Example~\ref{#1}}
\newrefformat{cond}{Condition~\ref{#1}}

\newcommand{\tp}{\intercal}

\usepackage[T1]{fontenc}
\usepackage[utf8]{inputenc}
\usepackage{authblk}
\title{Masked Transformer for  Electrocardiogram Classification}

\author[1,\#,*]{Ya Zhou}
\author[1,\#]{Xiaolin Diao}
\author[1]{Yanni Huo}
\author[1]{Yang Liu}
\author[2,*]{Xiaohan Fan}
\author[3,*,\dag]{Wei Zhao}
\affil[1]{Department of Information Center, Fuwai Hospital, Chinese Academy of Medical Sciences and Peking Union Medical College, Beijing, 100037, China}
\affil[2]{
	Cardiac Arrhythmia Center, Fuwai Hospital, National Center for Cardiovascular Diseases, Chinese Academy of Medical Sciences and Peking Union Medical College, Beijing, 
	China,  100037, China}
\affil[3]{Fuwai Hospital, National Center for Cardiovascular Diseases, Chinese Academy of Medical Sciences and Peking Union Medical College, Beijing, 100037, China}

\date{}

\begin{document}
\maketitle
\def\thefootnote{\#}\footnotetext{The authors contributed equally. }
\def\thefootnote{\arabic{footnote}}
\def\thefootnote{*}\footnotetext{Corresponding authors: Ya Zhou (zhouya@fuwai.com), Xiaohan Fan (fanxiaohan@fuwaihospital.org), Wei Zhao (zw@fuwai.com) }\def\thefootnote{\arabic{footnote}}
\def\thefootnote{ \dag}\footnotetext{Supervision: Wei Zhao}

\begin{abstract}
	Electrocardiogram (ECG) is one of the most important diagnostic tools in clinical applications. With the advent of advanced algorithms, various deep learning models have been adopted for ECG tasks.  However, the potential of Transformer for ECG data has not been fully realized, despite their widespread success in computer vision and natural language processing. In this work, we present Masked Transformer for ECG classification (MTECG), a simple yet effective method which significantly outperforms recent state-of-the-art algorithms in ECG classification.  
	Our approach  
	adapts the image-based masked autoencoders to self-supervised representation learning from ECG time series. We utilize a lightweight Transformer for the encoder and a 1-layer Transformer for the decoder. The ECG signal is split into a sequence of non-overlapping segments along the time dimension, and learnable positional embeddings are added to preserve the sequential information.  We construct the Fuwai dataset comprising 220,251 ECG recordings with a broad range of diagnoses,  annotated by medical experts, to explore the potential of Transformer. 
	A strong pre-training and fine-tuning recipe is proposed from the empirical study.  The experiments demonstrate that the proposed method increases the macro F1 scores by 3.4\%-27.5\% on the Fuwai dataset, 9.9\%-32.0\% on the PTB-XL dataset, and 9.4\%-39.1\% on a multicenter dataset, compared to the alternative methods. 
	We hope that this study could direct future research on the application of Transformer to more ECG tasks.
\end{abstract}

\section{Introduction}
\label{sec:intro}

The electrocardiogram (ECG) is one of the most commonly used non-invasive tools in clinical applications, recording the cardiac electrical activity and providing valuable diagnostic clues for systemic conditions  \citep{siontis2021artificial}.
With the  algorithmic advances in computer vision (CV) several years ago, 
the convolutional neural network (CNN) has 
been adapted to solve various ECG classification tasks, such as detecting arrhythmia \citep{hannun2019cardiologist, ribeiro2020automatic,  hughes2021performance}, left ventricular systolic dysfunction \citep{attia2019screening, attia2022prospective, yao2021artificial} and hypertrophic cardiomyopathy \citep{ko2020detection}. 

The field of CV has been transformed 
recently 
by the excellent performance of Transformer models \citep{vaswani2017attention, dosovitskiy2020image}. 
However, the progress of Transformer methods for ECG tasks lags behind those in CV. CNN remain the most common architecture for ECG data analysis \citep{siontis2021artificial, somani2021deep, wu2021scalable, pourbabaee2017deep}. A natural question is that can Transformer also be leveraged for ECG tasks.  
We attempt to 
answer the question in this paper.

One challenge is to adapt Transformer to effectively model ECG data. 
Commonly used ECG data are time-series signals with a unique temporal and spatial structure. For example, the standard 10 seconds 12-lead ECG signals with 500Hz sampling rate have 5000 time points. These signals typically consist of several beats with multiple waves, such as the P wave, QRS complex, T wave, and sometimes U wave \citep{sharma2021detection}.  Directly modeling each time step individually in Transformer might ruin the 
structural information 
and lead to heavy computation. 
	One potential solution is to split the signal into a sequence of non-overlapping segments. By treating these segments as tokens or image patches, one can readily utilize the design of Vision Transformer \citep[ViT, ][]{dosovitskiy2020image}. However, previous studies have demonstrated that the straightforward approach yields suboptimal classification accuracy \citep{li2021bat}. We believe that this is partly due to inappropriate training strategies. In this work, we will show this straightforward approach is capable of achieving excellent performance if appropriate training strategies are adopted.


Another challenge is the lack of available labeled datasets.  
It is known that the successful training of Transformer requires a large-scale dataset with labeled data, e.g., ImageNet-21k \citep{deng2009imagenet},  due to the lack of inductive bias \citep{dosovitskiy2020image}. However, creating labeled ECG datasets at scale is challenging since accurate annotation requires medical experts and considerable time. 
Although recent advancements have introduced larger ECG datasets like PTB-XL \citep{wagner2020ptb} and Chapman-Shaoxing \citep{zheng202012}, they still contain only a few tens of thousands of data points, making the Transformer model prone to overfitting \citep{li2021bat}. 
	To address this limitation, we build 
	the Fuwai dataset comprising 220,251 ECG recordings with a broad range of ECG diagnoses annotated by medical experts.


	Although the Fuwai dataset is much larger than publicly available ECG datasets, our experiments show that naively training a Transformer on it
	yields unsatisfactory results. We still observe overfitting, even when using a lightweight Transformer with 5.7 million parameters. 
	This problem might be caused by the information redundancy \citep{wei2019neural}, which typically exists in ECG time series. 
For example, each time point can be inferred from neighboring points, and some heartbeat cycles could be inferred from neighboring heartbeat cycles. 

In the field of natural images, the information redundancy 
	has
	been effectively addressed by masked modeling methods such as masked autoencoders \citep[MAE,][]{he2021masked} and SimMIM \citep{xie2022simmim}.

Inspired by their success, 
we propose a masked Transformer method for ECG classification, referred as MTECG. 
This method is a simple extension of the image-based MAE to self-supervised representation learning from ECG time series.
We split the ECG signal into a sequence of non-overlapping segments along the time dimension and 
adopt learnable positional embeddings to preserve the sequential order information and distinguish the segments. 
	To reduce the information redundancy, we utilize the lightweight Transformer with masked pre-training on unlabeled ECG data.
To encourage the encoder learning useful wave shape features, we adopt a 1-layer Transformer as the decoder 
and a fluctuated reconstruction target in the pre-training stage. 
To further reduce overfitting, we explore the regularization strategies such as layer-wise LR decay \citep{bao2021beit, clark2020electra} and DropPath \citep{huang2016deep} rate 
in the fine-tuning stage.

The main contributions of this work are summarized as follows.

1. \textbf{Novel Application}: 
We propose MTECG, a useful masked Transformer  method for ECG time series, as the extension of MAE originally proposed for the natural image analysis \citep{he2021masked}.  
	The derived lightweight model demonstrates the ability to classify a wide range of ECG diagnoses effectively, while remaining convenient for deployment in the clinical environment.

2. \textbf{New Dataset}: 
We build a comprehensive ECG dataset to evaluate various deep learning algorithms. The dataset consists of 220,251 recordings with 28 common ECG diagnoses annotated by medical experts and significantly surpasses the sample size of publicly available ECG datasets.

3. \textbf{Strong Pre-training and Fine-tuning Recipe}: We conduct comprehensive experiments to explore the training strategies on the proposed ECG dataset. The key components contributing to the proposed method are presented, including the masking ratio,  
training schedule length, fluctuated reconstruction target, layer-wise LR decay and DropPath rate.

4. \textbf{Excellent Performance}: We compare the proposed method with 
recent state-of-the-art algorithms
on a wide variety of tasks in both private and public ECG datasets. The experiments demonstrate that the proposed algorithm outperforms others significantly. Notably, it increases the macro F1 scores by 3.4\%-27.5\% on the Fuwai dataset, 9.9\%-32.0\% on the PTB-XL dataset, and 9.4\%-39.1\% on a multicenter dataset, compared to other methods.

The rest of this article is organized as follows. Section \ref{sec:related_work} briefly outlines the related works. Section \ref{sec:method} introduces the proposed method. 
The experiment setting, the properties and the practical performance of the proposed method can be founded in Section \ref{sec:exp}. We provide the additional discussion in Section \ref{sec:dis}. The conclusion of this work is summarized in Section \ref{sec:con}.

\section{Related work}
\label{sec:related_work}
\subsection{ECG classification}
ECG classification is one of the most important ECG data analysis tasks \citep{pyakillya2017deep, somani2021deep, siontis2021artificial}.  Traditionally, the classification for ECG signals is mainly based on expert features and conventional machine learning methods \citep{hong2020opportunities}. However, the algorithm performance is limited by data quality and expert knowledge \citep{hong2020opportunities}, precluding the usage in clinical applications \citep{ ribeiro2020automatic}.  To overcome these limitations,  numerous works have adapted end-to-end deep learning approaches to ECG tasks \citep{hannun2019cardiologist}. In comparison to the traditional methods, the deep learning approach can automatically learn useful features from raw ECG signals and achieve better performance in many classification tasks \citep{strodthoff2023ptb}.

\subsection{Transformer}

Transformer is an attention-based architecture 
which has rapidly become 
the dominant choice in natural language processing (NLP) since it was proposed \citep{vaswani2017attention, devlin2019bert, radford2018improving}. 
Due to its excellent performance, the field of computer vision (CV) has also been recently transformed by Transformer \citep{han2022survey, khan2022transformers}. 
Compared to CNN, Transformer can better utilize the global structure of the data and 
share shape recognition capabilities similar to those of the human visual system
\citep{naseer2021intriguing}. However, CNN still remains the most common architecture for ECG data analysis \citep{somani2021deep}. Although some works have 
	explored Transformer on ECG data analysis, 
	such as BaT \citep{li2021bat}, MaeFE \citep{zhang2022maefe} and CRT \citep{zhang2023self},  a comprehensive evaluation remains lacking and the potential of Transformer in this area is 
	not fully realized.
We will compare the performance of the proposed method with BaT, MaeFE and CRT.

\subsection{Self-supervised learning} 
Self-supervised learning is a popular method to utilize the unlabeled data and enhance the algorithm performance \citep{liu2021self}. 
Contrastive and  generative learning are two main self-supervised learning approaches \citep{krishnan2022self}. 

Contrastive learning 
	aims to model sample similarity and dissimilarity through a contrastive estimation objective \citep{gutmann2010noise, liu2021self}, and its performance  
	heavily relies on 
	the data augmentation strategy \citep{he2021masked, wang2023contrastive}. Various contrastive learning methods have been proposed for ECG data analysis, notable examples including CLECG \citep{chen2021clecg} and CLOCS \citep{kiyasseh2021clocs}. 
	A previous study suggests that CLECG generally outperforms CLOCS across multiple datasets \citep{chen2021clecg}.
	In this paper, we compare CLECG with our method.

Generative learning, such as GPT \citep{radford2018improving}, BERT\citep{devlin2019bert}, BEiT \citep{bao2021beit} and MAE \citep{he2021masked},  is another main self-supervised learning method. 
	Although steadily gaining attention in NLP and CV, 
	its effectiveness for ECG analysis tasks has yet to show significant performance improvements. Recent studies have explored the use of masked modeling techniques on ECG data, introducing models such as MaeFE  \citep{zhang2022maefe}  and CRT \citep{zhang2023self},  which have achieved superior results in ECG classification compared to other approaches. 
	Nonetheless,
	the full potential of masked modeling for ECG classification remains untapped. In this paper, we present a novel approach that significantly outperforms the state-of-the-art in this domain.

\section{Methodology}
\label{sec:method}
\begin{figure*}[!hbt]
	\centering
	\includegraphics[width = 150 mm 
	]{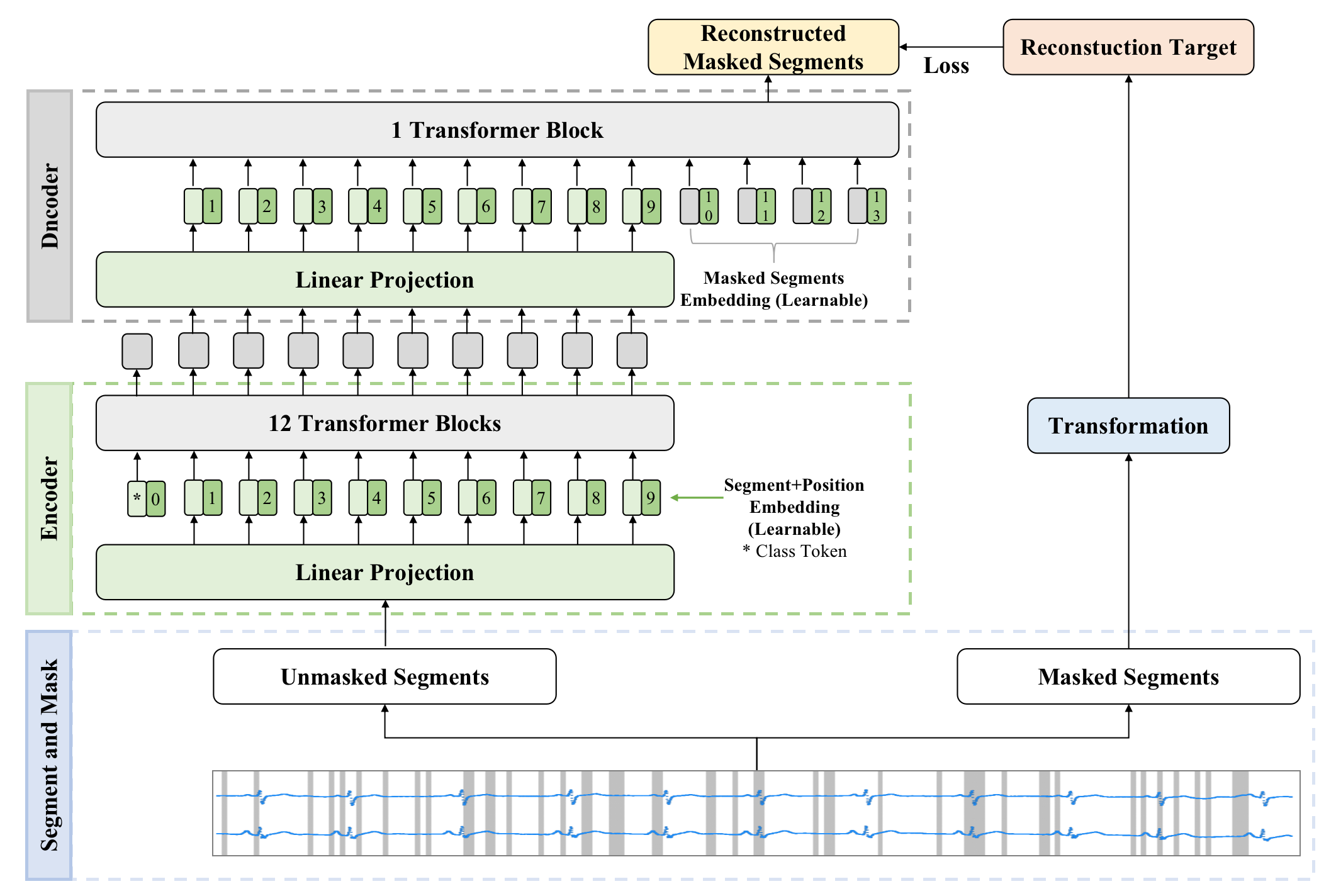}
	\caption{An example of the masked pre-training method. The original ECG signals are split to non-overlapping segments and a subset of these segments is masked out. The unmasked segments were used to reconstruct the fluctuated transformation of the masked segments through Transformer blocks. The lead and sequence information of the ECG signals are preserved by learnable positional embeddings. The masked segments are represented by learnable embeddings in the reconstruction task. 
		\label{pretrain_plots}}
\end{figure*}

Our approach expands image MAE \citep{he2021masked} to ECG time series and follows the pretrain-and-finetune  paradigm. The framework consists of 5 major components, i.e., segment operation, mask operation, encoder, decoder, and reconstruction target.  As shown in Figure \ref{pretrain_plots}, the pre-training stage involves all components.  As for fine-tuning, the decoder, reconstruction target and mask operation will be discarded. In the subsequent subsections, we detail each component and the training strategy.

\subsection{Segment operation}

For simplicity, this paper focuses on fixed-length ECG data. This type of ECG signal can be denoted as  $\mathbf X^\sharp = (X_{k,q}^\sharp) \in \mathbb{R}^{K \times Q}$,  where  $K$ is the number of leads and $Q$ is the length  of ECG. In many applications, $Q$ is much larger than $K$. For example, in the case of standard 12-lead ECGs, with a duration of 10 seconds and a sampling frequency of 500 Hz, we have $Q=5000$ and $K=12$.
When dealing with ECG signals of this length, directly modeling each time step individually using a Transformer would lead to heavy computation due to the quadratic cost 
of the self-attention operation. 
Moreover, ECG signals comprise multiple beats, each of which typically consists of various components such as the P wave, QRS complex, T wave, and U wave. If we were to model the signals at the individual time point level, we might ignore the wave shape information.

Alternatively, we split the ECG signal $ \mathbf X^\sharp$ into a sequence of non-overlapping segments $\mathcal{S} = \{\mathbf X_1, \ldots,  \mathbf X_T\}$, where $T$ is the sequence length.  The structural information of waves could be preserved in the segments and the sequence length could be remarkably reduced. 
Each segment $\mathbf X_t =(X_{k,q}^t)\in \mathbb{R}^{K \times (Q/T)}$ represents a subset of the original signal, where $\mathbf X_t  =(X_{k,q}^t)\in  \mathbb{R}^{K \times (Q/T)}  
$ and $X_{k,q}^t = X_{k, \{(t-1)*Q/T +q\}}^\sharp$ for $t=1,\ldots,T$, $k=1,\ldots,K, q=1,\ldots, Q/T$. These ECG segments can be regarded as image patches in computer vision, or tokens (words) in natural language processing. Similar to  image MAE \citep{he2021masked}, we utilize a linear projection to resize the vectorization of the ECG segments. 
However, unlike their work, we incorporate learnable positional embeddings to preserve the sequential order information and distinguish the segments, as shown in  \eqref{def:encoder} and \eqref{def:decoder}.

\subsection{Mask operation}
As mentioned in Section \ref{sec:intro}, we employ a masked pre-training method to address information redundancy and overfitting. 
Specifically, we employ a uniform random sampling technique to select segments from $\mathcal{S}$ without replacement, while the remaining segments are masked. 
For notational simplicity, we denote the unmasked segments as 
\[
\mathcal{S}_{unmask} = \{\mathbf  X_{i_1}, \ldots, \mathbf X_{i_{S}}  \}, 
\]
and the masked segments as
\[
\mathcal{S}_{mask} = \{\mathbf  X_{j_{1}}, \ldots, \mathbf X_{j_{S^\prime}}  \},
\]
Here, $i_s$ and $j_{s^\prime}$ are chosen from the set $\{1,\ldots,T\}$, with $s=1,\ldots,S$ and $s^\prime=1,\ldots,S^\prime$. The segments in $\mathcal{S}_{unmask}$ are sent to the encoder, while those in $\mathcal{S}_{mask}$ will be transformed into the reconstruction targets. 
Although the case $T=S + S^\prime$ is the main focus of this paper, the entire framework is also applicable to the case where $T< S + S^\prime$.

\subsection{Encoder}
\label{sec:enco}

In our settings, we can easily utilize the design of ViT \citep{dosovitskiy2020image}. 
We use a stack of standard  Transformer \citep{vaswani2017attention, dosovitskiy2020image} as the encoder. 
For the natural image, MAE \citep{he2021masked} incorporates vanilla positional embeddings (the sine-cosine version) to extract location information. 
However, previous studies demonstrate that this vanilla positional approach might be unable to fully exploit the important features of times series data \citep{wen2022transformers}. 
Therefore, we utilize learnable positional embeddings to effectively model the lead and sequence information in the ECG time series.

We briefly introduce the encoder below.  Denote the layer normalization \citep{ba2016layer}, the multi-headed self-attention and multi-linear perception blocks used in ViT \citep{dosovitskiy2020image} as $LN(\cdot)$ , $MSA(\cdot)$ and $MLP(\cdot)$, respectively. We use $ \mathbf x_{i_s}^\tp \in  \mathbb{R}^{KQ/T}$  to denote the vectorization of $\mathbf X_{i_s}$ for $s=1, \ldots, S$. Suppose $D$ is the latent vector size. Let the linear projection matrix $\mathbf  E \in \mathbb{R}^{ (KQ/T) \times D }$, the auxiliary token $ \mathbf x_{class}^\tp  \in \mathbb{R}^{D}$ , and the positional embedding $\mathbf e_{pos} =(\mathbf e_{0}, \mathbf e_1  , \cdots, \mathbf e_T )^\tp   \in \mathbb{R}^{D(T+1) }$ be  learnable. Here $\mathbf e_t^\tp  \in \mathbb{R}^D, t=0, 1,\ldots, T$ are used to preserve the sequential order information and distinguish the segments. During pre-training, only the unmasked segments are fed to the encoder, which can be written as 
\begin{equation}
	\label{def:encoder}
	\begin{aligned}
		\mathbf z_0 &= [\mathbf x_{class} ; \mathbf x_{i_1} \mathbf E , \cdots, \mathbf x_{i_S} \mathbf E  ]  + [\mathbf e_{0}; \mathbf e_{i_1}; \ldots; \mathbf e_{i_S}  ], \\
		\mathbf z_{l}^\prime & = MSA(LN(\mathbf z_{l-1})) + \mathbf z_{l-1}, l=1,\ldots,L, \\
		\mathbf z_{l}& = MLP(LN(\mathbf z_{l}^\prime)) + \mathbf z_{l}^\prime, l=1,\ldots,L, \\
	\end{aligned}
\end{equation}
where 
$L$ is the  number of blocks. 
For simplicity, we denote the output $\mathbf z_L$ as 
\begin{equation*}
	\label{def:encoder_output}
	\mathbf z_L = [\mathbf z_L^0; \mathbf z^{i_1}_L , \cdots ; \mathbf z^{i_S}_L ],
\end{equation*}
where 
the encoded auxiliary token $(\mathbf z^0_L)^\tp \in  \mathbb{R}^{D}$ is discarded and the encoded unmasked segments $(\mathbf z^{i_s}_L)^\tp \in \mathbb{R}^{D}, s=1,\ldots, S$ are used for the reconstruction targets in pre-training. In contrast, during fine-tuning, the encoder is applied to all segments and both the encoded auxiliary token and segments could be used for the downstream task.

\subsection{Decoder}
\label{sec:dec} 
To encourage the encoder learning useful wave shape features and reduce the computation cost, we use a  1-layer Transformer as the decoder. 
Suppose $D^\prime$ is the latent vector size. Let $ \mathbf  E^\prime \in \mathbb{R}^{D \times D^\prime }$,  $\mathbf E_0 \in \mathbb{R}^{D^\prime \times (KQ/T)} $, $\mathbf e_{m}^\tp  \in \mathbb{R}^{D^\prime}$ and 
$\mathbf e_{pos}^\prime = (\mathbf e_1^\prime  , \cdots, \mathbf e_T^\prime )^\tp  \in \mathbb{R}^{T D^\prime} $ 
be learnable components. Here $ ( \mathbf e_t^{\prime} )^\tp \in \mathbb R^{D^\prime}$ for  $t=1,\ldots,T$ serve as the positional embeddings and $\mathbf e_m$ is used to denote the masked segments.  The decoder can be written as 
\begin{equation}
	\label{def:decoder}
	\begin{aligned}
		\tilde{\mathbf z}_0 &= [\mathbf z^{i_1}_L \mathbf E^\prime ; \cdots; \mathbf z_L^{i_S} \mathbf E^\prime; \mathbf e_m; \cdots; \mathbf e_m ]   \\ & \quad \quad \quad + [\mathbf e_{i_1}^\prime; \ldots; \mathbf e_{i_S}^\prime; \mathbf e_{j_1}^\prime; \cdots; \mathbf e_{j_{S^\prime}}^\prime  ]\\
		\tilde{\mathbf z}^\prime_1 & = MSA(LN( \tilde{\mathbf z}_{0})) + \tilde{\mathbf z}_{0}, \\
		\tilde{\mathbf z}_1 & = MLP(LN(\mathbf z^\prime_1)) + \mathbf z^\prime_1, \\
	\end{aligned}
\end{equation}
where $	\tilde{\mathbf z}_{1}^\prime $ can be written as 
$ \tilde{\mathbf z}_{1}^\prime = [\tilde{\mathbf z}^{i_1}_1 ; \cdots ; \tilde{\mathbf z}^{i_S}_1; \tilde{\mathbf z}^{j_1}_1 ; \cdots ; \tilde{\mathbf z}^{j_{S^\prime }}_1 ]$. Here 
$(\tilde{\mathbf z}_1^{i_s})^\tp, (\tilde{\mathbf z}_1^{j_{s^\prime}} )^\tp  \in \mathbb{R}^{ D^\prime}$ for $s=1,\ldots,S$, $s^\prime=1,\ldots,S^\prime$ and  $\tilde{\mathbf z}_1^{j_{s^\prime}} $ will be used to generate the output  $\tilde{ \mathbf x}_{j_{s^\prime}}$ though
\begin{equation}
	\label{def:decoder_output2}
	\tilde{ \mathbf x}_{j_{s^\prime}} = \tilde{\mathbf z}_1^{j_{s^\prime}} \mathbf E_0, s^\prime =1,\ldots, S^\prime. 
\end{equation}

\subsection{Reconstruction target} 
\label{seq:re_target}
The art of masked modeling lies in the choice of pretext tasks \citep{liu2021self}. Due to the cyclic nature of the ECG signals, directly reconstructing the original signal might weaken the learning of wave shape features. To encourage the encoder learning useful features, we adopt a fluctuated reconstruction target in pre-training. 
To be more specific, 
we denote $f: \mathbb{R}^{QW/T} \to \mathbb{R}^{QW/T} $ as 
a pre-defined fluctuated mapping.  
We utilize the following mean square error (MSE) as the pre-training loss function
\begin{equation*}
	\label{def:mse}
	MSE =  \sum_{s^\prime=1}^{S^\prime }\Vert \tilde{\mathbf x}_{j_{s^\prime}} - f(\mathbf x_{j_s^\prime } ) \Vert_2^2.
\end{equation*}
Here $\tilde{\mathbf x}_{j_{s^\prime}} $ is defined in \eqref{def:decoder_output2}, $\mathbf x_{j_s^\prime } = (x_{j_s^\prime, j})_{j=1}^{KQ/T}$ 
is the vectorization of $\mathbf X_{j_{s^\prime}}$, where $\mathbf X_{j_{s^\prime}} \in \mathcal{S}_{mask}$ is a masked segment. 
In this paper, we explore two options for the fluctuated mapping $f$. 
The first one is per-segment normalization and the  $j$-th element of $f(\mathbf x_{j_s^\prime } )$ is defined as 
\begin{equation}
	\label{eqn:per-segment}
	f(\mathbf x_{j_s^\prime })_j = \frac{x_{j_s^\prime, j} - \mu_{j_s^\prime}  }{\sqrt{\sigma^2_{j_s^\prime} + \epsilon}},  
\end{equation}
where 
\[
\mu_{j_s^\prime} = \frac{T}{KQ} \sum_{j=1}^{KQ/T} x_{j_s^\prime, j}, \quad  \sigma_{j_s^\prime }^2 = \frac{T}{KQ} \sum_{j=1}^{KQ/T} (x_{j_s^\prime, j} -  \mu_{j_s^\prime}  ),
\]
and $\epsilon$  is a constant used for the numerical stability. 
The second choice is just a a simple squaring operation. We define
the $j$-th element of $f(\mathbf x_{j_s^\prime } )$ as 

\begin{equation}
	\label{eqn:squaring}
	f( \mathbf x_{j_s^\prime})_j = sign( x_{j_s^\prime, j} ) \vert x_{j_s^\prime, j} \vert ^{0.5}.  
\end{equation}
Both of these options have been demonstrated to enhance local contrast in time-series ECG data. The per-segment normalization \eqref{eqn:per-segment} is the counterpart of the per-patch normalization employed in MAE \citep{he2021masked}, whereas the squaring operation \eqref{eqn:squaring} is a simple and effective approach proposed in this study.

\subsection{Training strategy}
Our approach follows the the pretrain-and-finetune paradigm. 
Firstly, we pre-train the model to learn a useful  ECG representation through masked modeling, as illustrated in Figure \ref{pretrain_plots}. Secondly, we follow MAE \citep{he2021masked} and fine-tune the pre-trained encoder with an added classification head for the downstream tasks.  
Since ECG signals have a distinct nature compared to images and language, directly applying the training recipe of images or language might lead to poor results. 
We investigate several components such as masking ratio and reconstruction targets to effectively train the model. Moreover, considering the heavier redundancy in ECG signals,  we also dive into the regularization hyper-parameters, 
such as layer-wise LR decay \citep{bao2021beit, clark2020electra} and DropPath \citep{huang2016deep} rate, 
during fine-tuning to further reduce overfitting.

\section{Experiments}
\label{sec:exp}
In this section, we explore the property of the proposed method based on the Fuwai dataset consisting 220,251 ECG recordings with a broad range of diagnoses annotated by medical experts. Additionally, we evaluate the prediction performance of the proposed method on both private and public ECG datasets. 

\subsection{Datasets}
The \textbf{Fuwai dataset} consists of 220,251 ECG recordings from 173,951 adult patients in Fuwai Hospital of Chinese Academy of Medical Sciences. This dataset encompasses 28 different diagnoses, and the diagnoses for each ECG recording were annotated by two certified ECG physicians. The dataset was split into three sets based on patients, with a ratio of 8:1:1 for the training, validation, and testing sets, respectively. This study was approved by the Ethics Committee at Fuwai Hospital, and no identifiable personal data of patients were used.

The \textbf{PCinC dataset} was collected from the  PhysioNet/Computing in Cardiology Challenge 2021 \citep{reyna2021will}. The public portion of the challenge datasets contains 88,253 12-lead ECG recordings gathered 
from 8 datasets, 
including PTB \citep{bousseljot1995nutzung}, 
PTB-XL \citep{wagner2020ptb}, Chapman-Shaoxing \citep{zheng202012}, Ningbo \citep{zheng2020optimal}, CPSC \citep{liu2018open}, CPSC-Extra \citep{liu2018open}, INCART \citep{tihonenko2007st}, and G12EC\citep{reyna2021will}. The diagnoses of these ECG datasets were encoded to 133 diagnoses using approximate Systematized Nomenclature of Medicine Clinical Terms (SNOMED-CT) codes, with 30 diagnoses were selected as being of clinical interest \citep{reyna2022issues}. 
Following the competition, we  merged 8 diagnoses into 4 labels. We then consider subjects with complete 12-lead signals, sampled at a rate of 500 Hz over a duration of 10 seconds, resulting in a total of 79,574 ECG recordings. Furthermore, we refine the dataset by selecting only labels with an incidence rate of at least 0.5\%, further reducing the number of labels to 25. To evaluate the algorithms, we randomly divided the entire dataset into training, validation, and testing sets in a ratio of 7:1:2, respectively.

The \textbf{PTB-XL dataset} is a commonly used  dataset to evaluation ECG classification algorithms \citep{strodthoff2020deep}. It is 
a part of 
the aforementioned 
challenge datasets
and includes 
21,837
ECG recordings. Consistently, we utilize SNOMED-CT labels with an incidence of at least 0.5\% and remove the sample without complete 12-lead signals, resulting in the consideration of 21,836 recordings with 22 labels.
We follow the suggested 10-fold split by
\cite{strodthoff2020deep}, where folds 1-8, 9 and 10 are training, validation, testing sets, respectively.

\subsection{Metrics}
Our clinical objective is to detect the majority of positive cases while minimizing the occurrence of false alarms. 
In binary classification, the F1 score is a commonly used 
performance metric that takes this trade-off into account. It is calculated as the harmonic mean of sensitivity (recall) and precision.
When dealing with multi-label classification in imbalanced data, such as our ECG datasets, 
the macro metrics can provide a better assessment of the overall classification performance \citep{bagui2021resampling}. 
Therefore, we use the macro F1 score as the major evaluation metric throughout this paper. 




\subsection{Default setting}
We adopt MTECG-T as the default backbone, a lightweight Transformer architecture, 
which will be introduced in Section \ref{subsec:scaling_exp}. Following \cite{li2021bat}, we use 25 as the default segment size, which yields the sequence length $T=200$ in our experiments. In addition, we use a mask ratio of 25\%, a per-segment normalization reconstruction target, a light weight decoder with a latent dimension $D^\prime=128$ alongside a self-attention head of 4, and a global pooling classifier as the default setting. 

Initially, we pre-train the model on the training set. 
During pre-training, we utilize an AdamW optimizer with a cosine learning rate schedule. The default hyperparameters include a momentum of $(\beta_1, \beta_2)=(0.9, 0.95)$, a weight decay of 0.05, a learning rate of 0.001, a batch size of 256, and a total of 1600 training epochs with a warm-up epoch of 40.

After pre-training, we fine-tune the pre-trained encoder on the same dataset. For fine-tuning, we also use the AdamW optimizer and the cosine learning rate schedule. 
The default hyperparameters includes 
a momentum of $(\beta_1, \beta_2)=(0.9, 0.999)$, a weight decay of 0.05, a learning rate of 0.001, a batch size of 256, a DropPath rate of 0.4, a layer decay rate of 0.6, and a warm-up epoch of 5. We train the model for 50 epochs using the binary cross entropy loss and select the one corresponding to the highest macro F1 scores on the validation set as our final model. 

\begin{figure}[!hbt]
	\centering
	\includegraphics[width = 75 mm 
	]{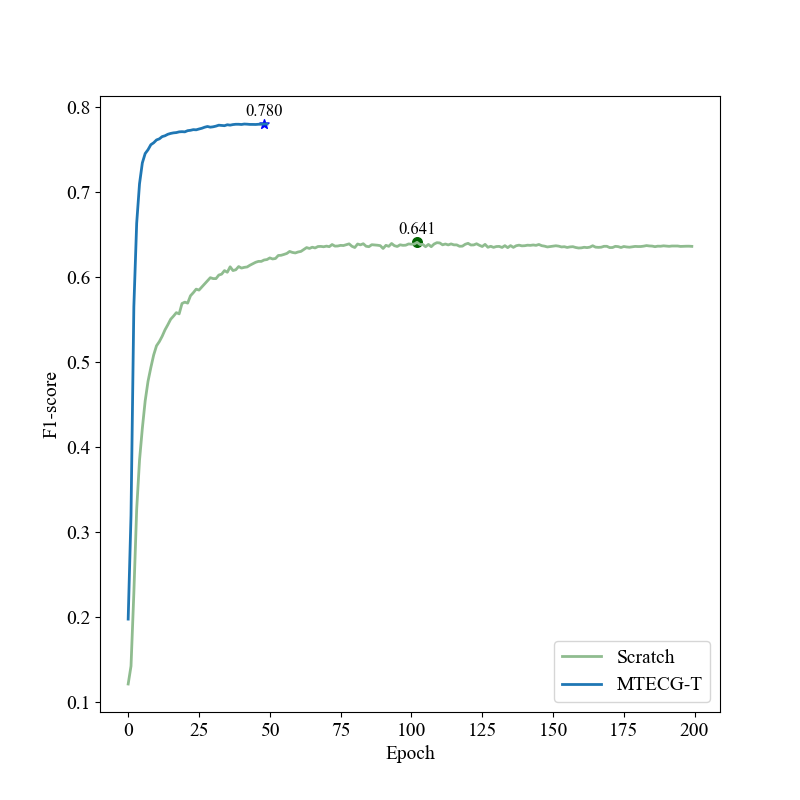}
	\caption{Performance comparison between masked pre-training and training from scratch on the Fuwai dataset. The optimal epoch for masked pre-training is found to be 48, whereas for training from scratch, it is 102.
	\label{com_pt_scratch} }
\end{figure}

For comparison, 
we also train the model from scratch for 200 epochs using the same recipe as the fine-tuning process. Figure \ref{com_pt_scratch} depicts the comparison result, from which we can see the masked pre-training significantly improve the classification performance.

\subsection{Ablation study}
In the ablation study, we explore the properties of important components for the proposed method on the Fuwai dataset, and report the marco F1 score on the validation set.

\subsubsection{Masking ratio}
\begin{figure*}[!hbt]
	\centering
	\includegraphics[width = 75 mm
	]{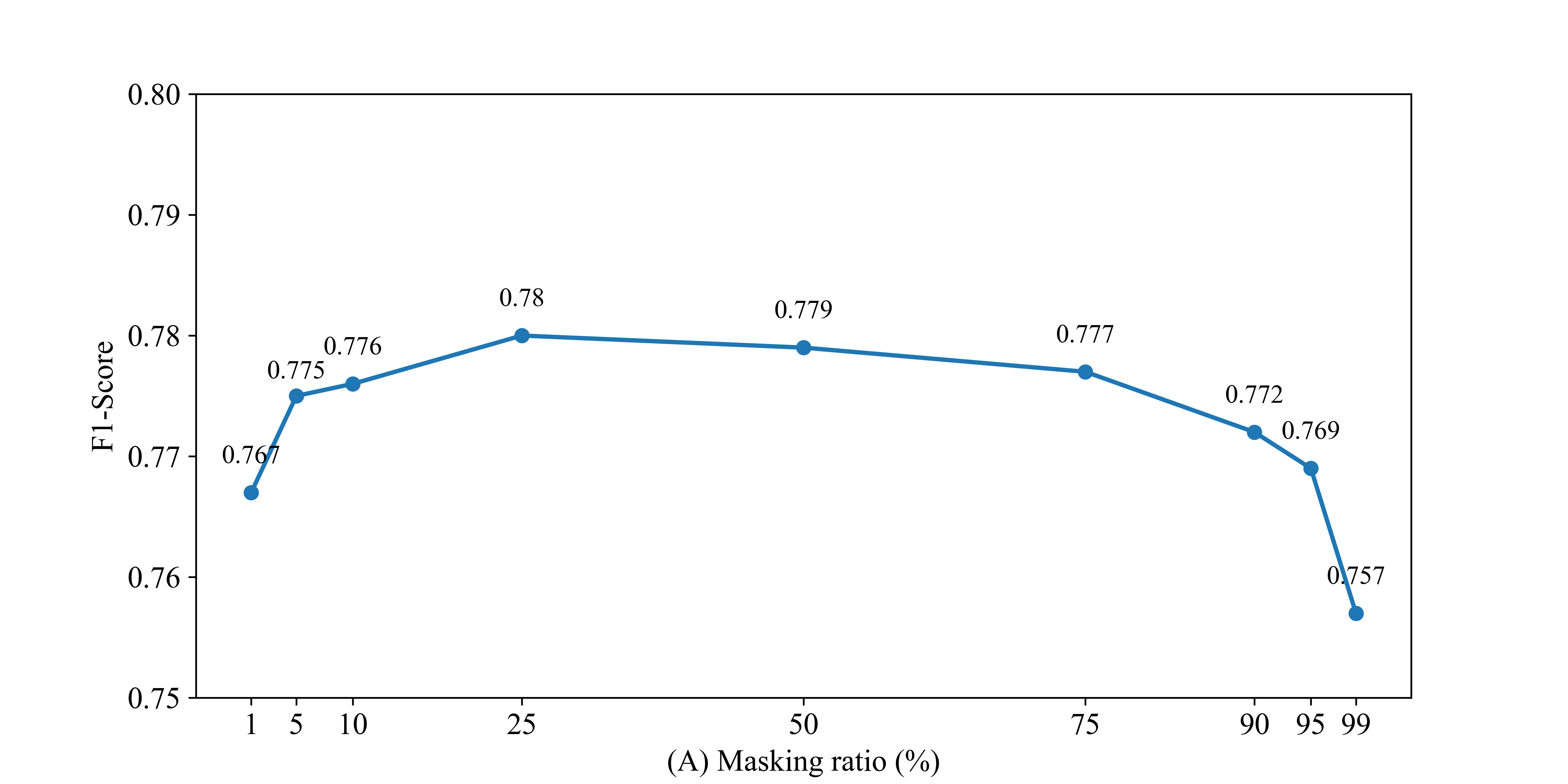}
	\includegraphics[width = 75 mm
	]{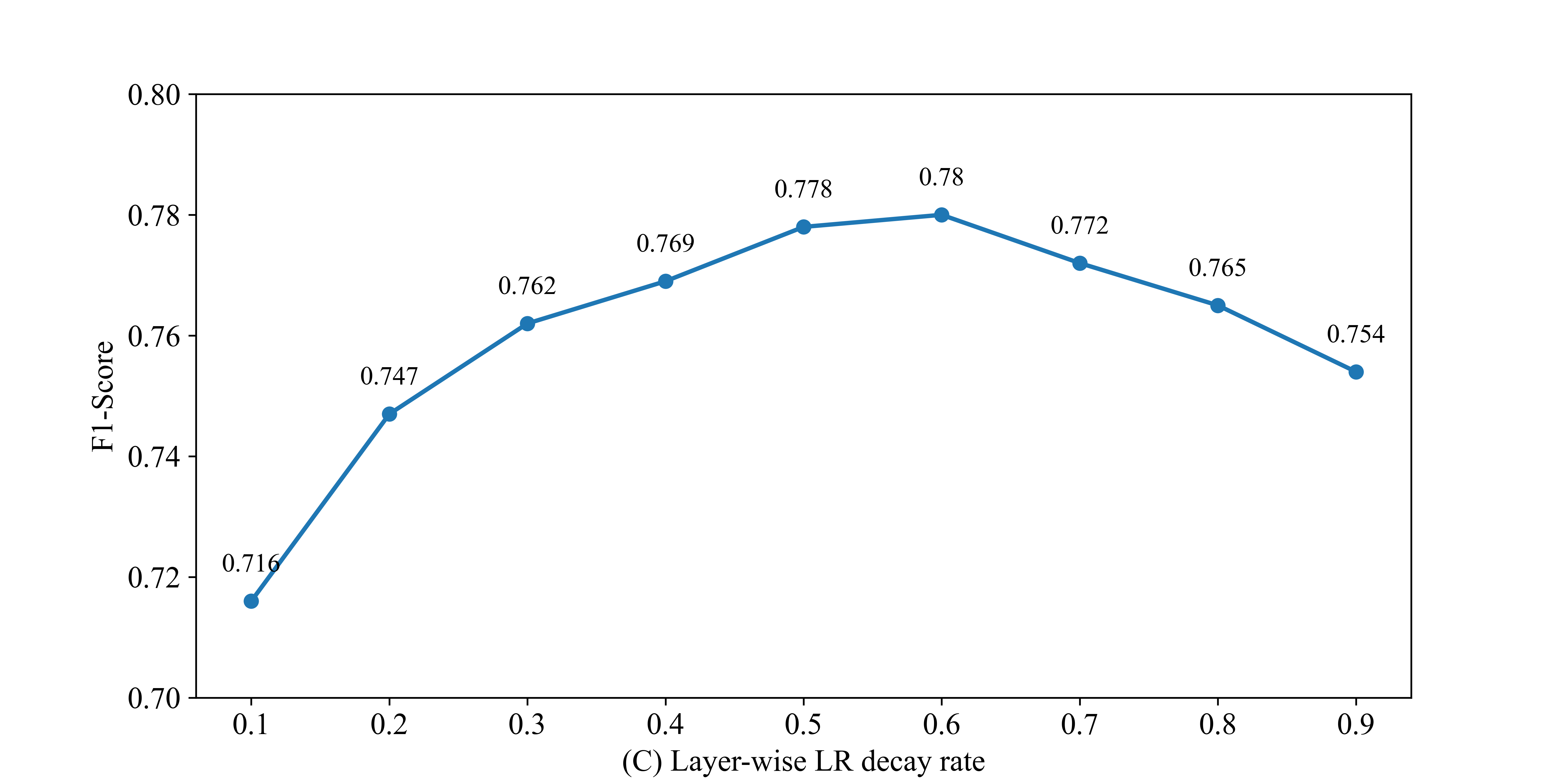}
	\includegraphics[width = 75 mm
	]{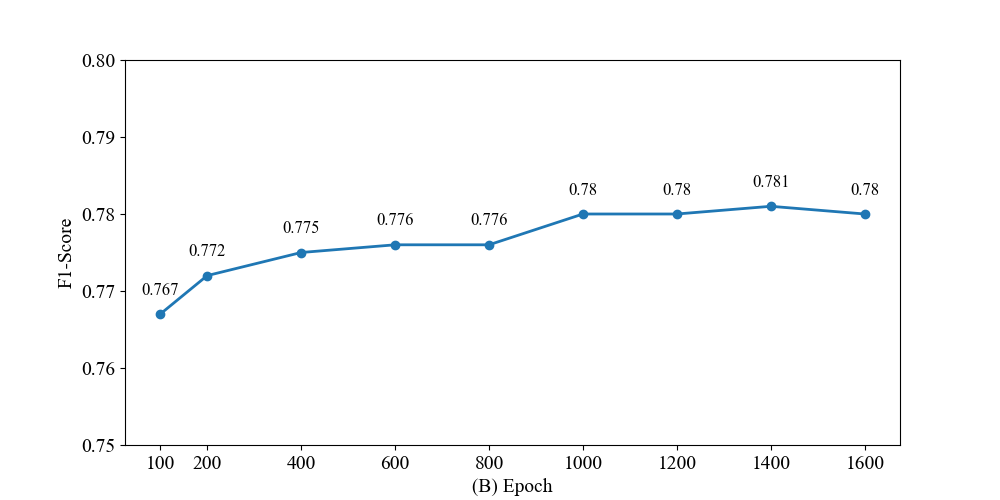}
	\includegraphics[width = 75 mm
	]{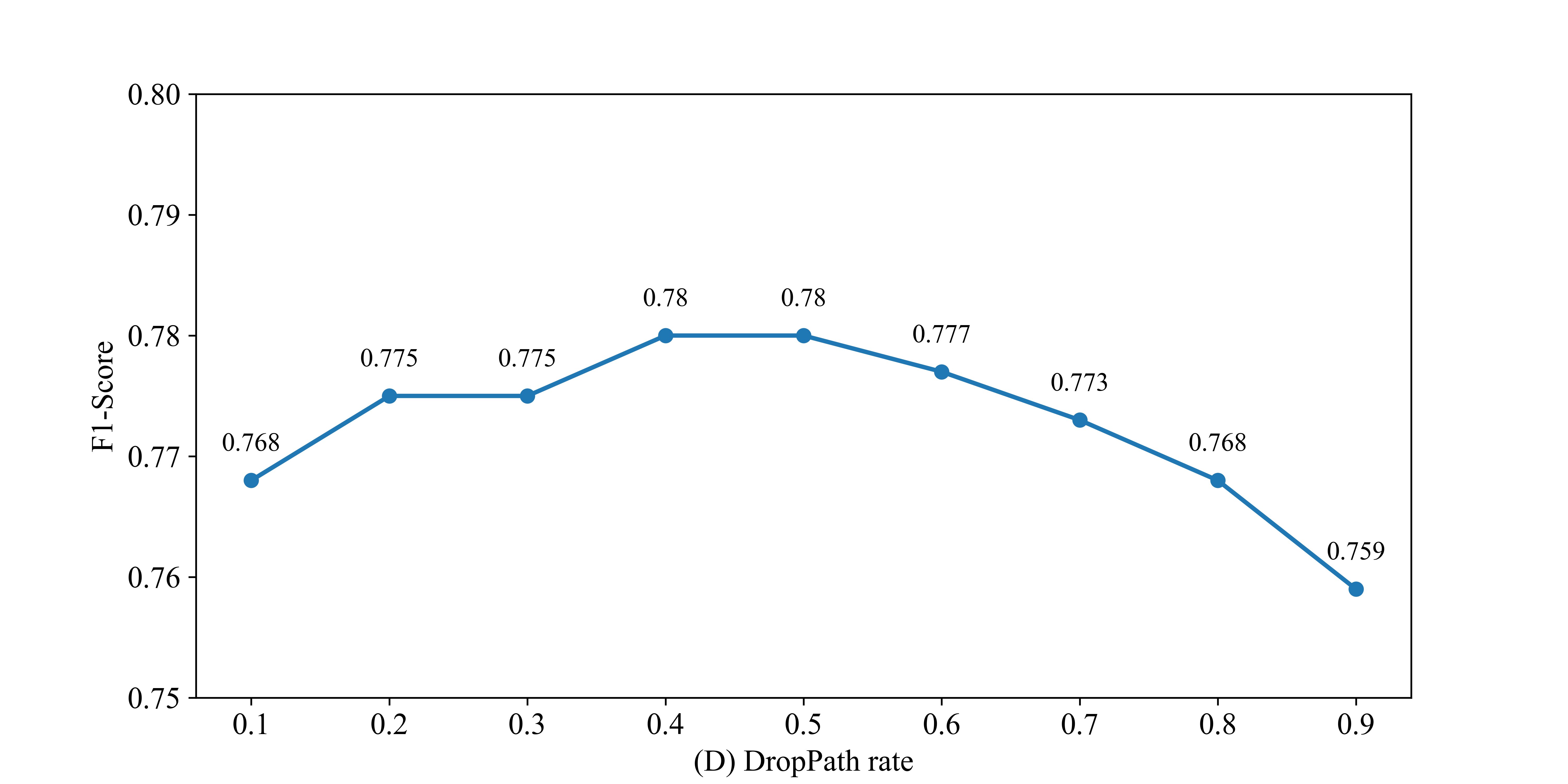}
	
	\caption{Classification performance in the ablation study. 
		The first column, from top to bottom, corresponds to masking ratio and pre-training schedule lengths, respectively. Similarly, the second column, from top to bottom, corresponds to layer-wise LR decay and DropPath rate, respectively.
		\label{recipe_plot} }
\end{figure*}

The validation marco F1 scores of the proposed MTECG-T under multiple masking ratios are summarized in Figure \ref{recipe_plot} (A). 
The optimal masking ratio of the proposed method is 25\% and it performs stably well across a broad range of ratios  (5\%-75\%).
Interestingly, we observed that even at the extreme ratios of 1\% and 99\%, our proposed method significantly outperforms the model trained from scratch. These findings highlight the effectiveness of the masked pre-training.

\subsubsection{Training schedule}
We investigate the influence of the pre-training schedule length on the fine-tuning performance. 
Figure \ref{recipe_plot} (B) presents our findings. The performance improves with longer training and begins to saturate at 1000 epochs. 
Interestingly, 
even with shorter training schedules that had not yet reached saturation, such as the 100-epoch pre-training case, the proposed method exhibits significantly superior performance compared to the model trained from scratch.

\subsubsection{Reconstruction target}

\begin{figure*}[!hbt]
	\centering
	\includegraphics[width = 150 mm
	]{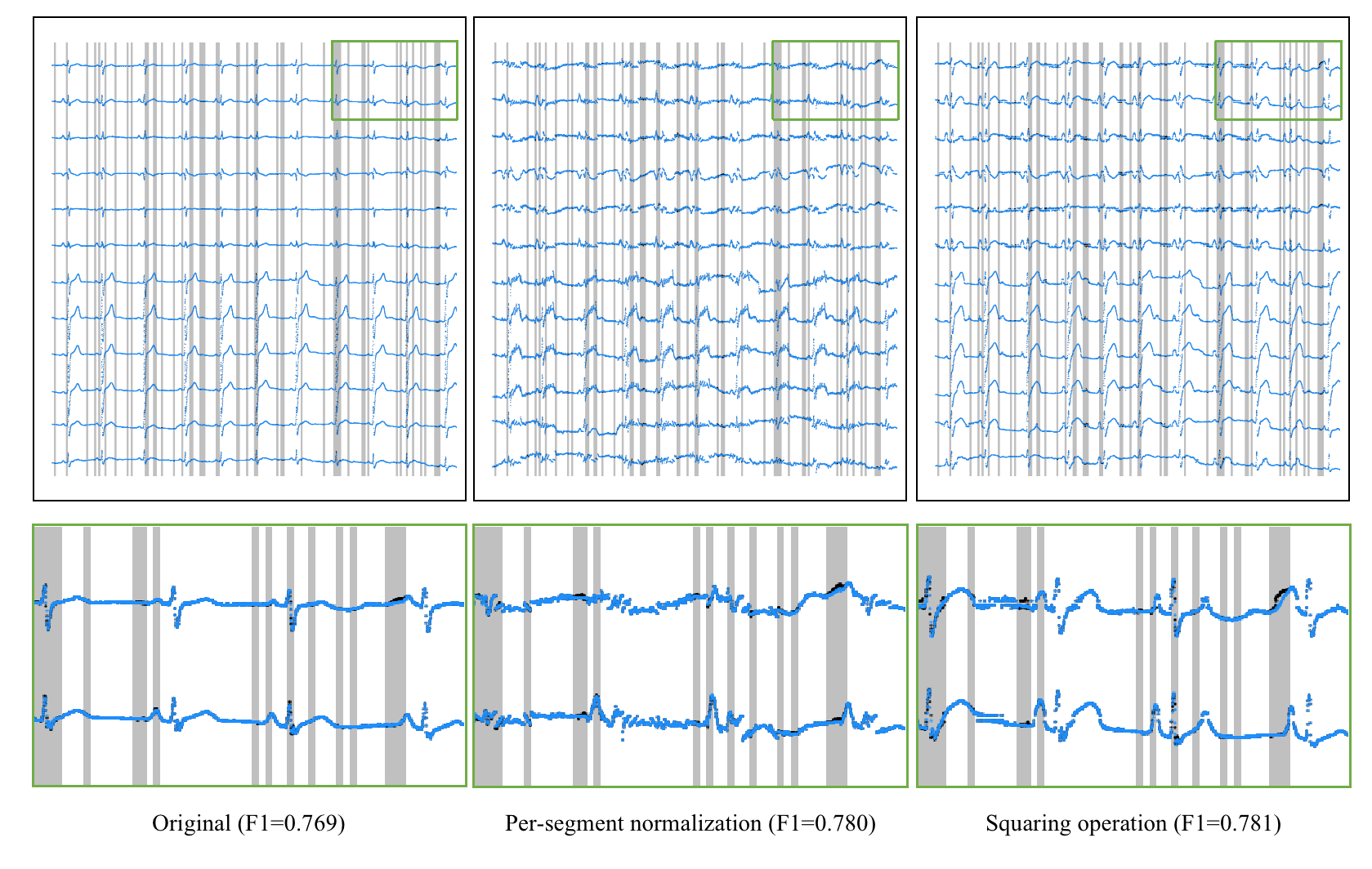}
	\caption{
		Reconstruction examples from one patient in the validation set under different reconstruction targets. The first row corresponds to the whole ECG signal, while the second row corresponds to a sequence of signal segments. The columns represent the original, per-segment normalization \eqref{eqn:per-segment}, and squaring operation \eqref{eqn:squaring} targets, respectively. The gray regions indicate the masked segments. The black point represents the original signal, while the blue point within the masked region represents the reconstructed signal.
		\label{reconstruct_plots} }
\end{figure*}

In this subsection, we compare different reconstruction targets. 
Figure \ref{reconstruct_plots} presents the reconstruction examples from one patient in the validation set and the corresponding fine-tuning performance. 
Our experimental results clearly demonstrate the beneficial impact of two transformation operations, i.e., per-segment normalization and squaring operation. We observe that both operations fluctuate the original signals, which might enhance the local contrast and amplify subtle changes. 

\subsubsection{Fine-tuning strategy}
We investigate the effects of regularization hyper-parameters, layer-wise LR decay \citep{bao2021beit, clark2020electra} and DropPath \citep{huang2016deep} rate on the fine-tuning performance. 
The results are presented in Figure \ref{recipe_plot} (C and D)
, which show the performance across a range of hyper-parameter values from 
0.1 to 0.9.
The layer-wise LR decay rate of 0.6 is identified as the optimal setting, while for the DropPath rate, options of 0.4 and 0.5 are considered. However, for simplicity, a DropPath rate of 0.4 is utilized later on.
Interestingly, we observe that the performance is slightly more sensitive to these hyper-parameters than to the masking ratio.

Furthermore, we conducted experiments on the encoded auxiliary token and segments for classification and 
found that their performances are comparable. 
Specifically, when using all encoded segments for classification through global pooling, the macro validation F1 score is 0.780, 
while using the auxiliary token directly yields a macro F1 score of 0.779.

\subsection{Scaling experiments}
\label{subsec:scaling_exp}

Noting that the model capacity matters to the generalization ability, we adopt Transformer of different model size for experiments, including MTECG-A (Atomic), MTECG-M (Molecular), MTECG-T(Tiny), MTECG-S (Small) and MTECG-B (Base), for which we change the number of heads $h$ for the encoder, keeping  $D/h=64$ and the layers $L=12$. Table \ref{ViT_archs} summarizes the architectures and their fine-tuning performance on the validation set of the Fuwai dataset. 
The fine-tuning performance initially increase and then fall as the model size increases.

\begin{table}[!htbp]
	\centering
	\caption{Variants of our  MTECG architecture and their fine-tuning performance.  
		The \#params column represents
		the number of trainable parameters. 
		\label{ViT_archs}\vspace{0.5ex} } 
	
	\fbox{
		\begin{tabular}{c|cccccc}  
			\multirow{2}*{Model}&	\multirow{2}*{Segment  size} & \multirow{2}*{$D$}& \multirow{2}*{
				$h$
			}  & \multirow{2}*{
				$L$
			}   & \multirow{2}*{\#params} & \multirow{2}*{F1}   \\ 
			\\  \hline
			MTECG-A &25&64 & 1  &12   & 0.9M&  0.752  \\
			MTECG-M &25& 128&   2  &  12 &   2.7M & 0.775  \\
			MTECG-T	&25& 192&   3 &  12 &    5.7M& \textbf{0.780} \\
			MTECG-S	&25&384&     6  &  12   &   21. 8M& 0.775 \\
			MTECG-B  &25& 768&     12  &  12   &   85.8M& 0.776 \\
		\end{tabular}  
	}
\end{table}  

\subsection{Comparison with the state-of-the-art algorithms}
	We compare our proposed MTECG-T with four state-of-the-art approaches: (i) CLECG \citep{chen2021clecg}, (ii) MaeFE \citep{zhang2022maefe}, (iii) CRT \citep{zhang2023self}, and (iv) BaT \citep{li2021bat}. These methods represent distinct techniques in the domain of ECG classification. Specifically, CLECG utilizes contrastive learning to enhance CNN performance. MaeFE and CRT employ masked pre-training for Transformer, which differs from our method in terms of model architecture, training strategies, and data preprocessing. BaT is specifically designed with a Transformer architecture to learn information from ECG heartbeats.	
	
	We conduct experiments across three different settings, indicated as Fuwai, PTB-XL, and PCinC in Table \ref{com_SOTA}. For the two-stage methods, including CLECG, MaeFE, CRT and MTECG-T, we develop algorithms as follow. In the first setting, we pre-train and fine-tune the models on the training set of the Fuwai dataset. In the second setting, we pre-train the models on the PCinC dataset, excluding PTB-XL, and then fine-tune them on the training set of PTB-XL. In the third setting, both pre-training and fine-tuning are performed on the training set of the PCinC dataset. For the single-stage method, i.e., BaT, we train the model from scratch on the training set of Fuwai, PTB-XL and PCinC, respectively.
	For these methods involved in the comparison, we adopt the data preprocessing, training strategies and hyper-parameters recommended in the corresponding reference papers for each method
		\citep{chen2021clecg, zhang2022maefe, li2021bat, strodthoff2020deep, zhang2023self}. Particularly, for CLECG, we adopt xresnet1d101 \citep{strodthoff2020deep} as the backbone. In the case of MaeFE, we employ wavelet transform filtering for data preprocessing, as proposed by \citep{martis2013ecg}, and 
		utilize the MTAE masking strategy with the ViT backbone, 
		as suggested in \citep{zhang2022maefe}. For CRT, we adopt the training strategy and backbone as detailed in \citep{zhang2023self}. For BaT, we 
		use the segmentation method \citep{makowski2021neurokit2} to obtain ECG heartbeats. 
	The algorithm comparison is demonstrated in Table \ref{com_SOTA}, from which we can see 
	MTECG-T significantly outperforms the alternatives across all datasets. Notably, MTECG-T increases the macro F1 scores by 3.4\%-27.5\% in Fuwai, 9.9\%-32.0\% in PTB-XL, and 9.4\%-39.1\% in PCinC, compared to the alternative methods.

\begin{table}[!htbp]
	\centering
	\caption{Prediction performance on 3 real datasets. Reported are the macro F1 scores on the testing set of each dataset. 
		\label{com_SOTA}\vspace{0.5ex} } 
	\fbox{
		\begin{tabular}{c|cccccc}  
			Methods & Fuwai & PTB-XL & PCinC   \\ 
			\hline
			CLECG & 0.740 & 0.518 & 0.590 \\
			\citep{chen2021clecg} & & & \\
			MaeFE & 0.705 & 0.533 & 0.605 \\
			\citep{zhang2022maefe} & & & \\
			CRT & 0.600 & 0.444 & 0.476 \\
			\citep{zhang2023self} & & & \\
			BaT & 0.702 & 0.462 & 0.561 \\
			\citep{li2021bat} & & & \\
			MTECG-T & \textbf{0.765} & \textbf{0.586} & \textbf{0.662} \\
		\end{tabular}  
	}
\end{table}

	
	\section{Discussion}
	\label{sec:dis}

	The application of deep learning for ECG classification has gained significant popularity in various domains.
	In this study, we extensively evaluate and explore the application of the innovative Transformer architecture and masked modeling methods, leveraging both private and publicly available ECG datasets. 
	Our experiments demonstrate promising performance of the proposed method, which also underscore their potential effectiveness in clinical applications. Moreover, our findings raise several questions as follows.

	1. \textbf{A standard ECG is worth 200 words?} In this paper,  
	the segments, architectures and training strategies are similar to techniques used in NLP.  One may wonder whether the deep learning methods for ECG will also follow a similar trajectory as NLP. 
	It is important to acknowledge that the simple self-supervised learning method in NLP enables benefits through model scaling. However, our scaling experiments demonstrate that the masked pre-training method fails to sustain scalable benefits, with performance exhibiting an initial rise followed by a subsequent decline as the model size increases. 
	One potential explanation could be the insufficient size of the training data. 
	Although larger than almost all publicly available ECG datasets, our dataset is still relatively small compared to the datasets commonly used in NLP and CV pre-training tasks. 
	
	%

	2. \textbf{The Transformer is suitable for ECG time series in lightweight regimes?} On the other hand, our experiments show that the lightweight Transformer could achieve excellent performance in ECG classification if proper training strategies are adopted. Note that our networks stem from the vanilla ViT. 
	The findings break the popular conception that the vanilla Transformer is not suitable for classification tasks in lightweight regimes, which is consistent with the observation of \cite{wang2023closer}. 
	These facts also suggest that employing appropriate pre-training techniques
	the naive network architectures 
	could be better than 
	specifically designed ones. 
	It is worth to mention the naive lightweight model are friendly to deployment in the clinical environment. 

	3. \textbf{ How important the training recipe is in the masked modeling?} 
	Previous emphasis was primarily on the masking strategy, training schedule length, and reconstruction target. However, in this work, we demonstrate that the fine-tuning performance is not only sensitive to the those components, but also to other components such as the layer-wise LR decay and DropPath rate. 
	These findings indicate that the successful application of masked modeling in domain-specific tasks might rely 
	on the implementation of proper pre-training and fine-tuning recipes.

	
	4. \textbf{Transformer or CNN for ECG classification?}
	Transformer has shown remarkable performance in various visual benchmarks, often matching or surpassing that of CNN. One might be curious about whether Transformer could potentially replace CNN for ECG classification tasks. 
	We compare the Transformer-based model pre-trained by the proposed masked method with the benchmark CNN-based model (xresnet1d101 \citep{strodthoff2020deep}) pre-trained by the contrastive method (CLECG), and observe that the Transformer-based model achieves significantly better performance.
	However, we cannot conclude that Transformer is a better architecture than CNN for ECG classification due to the different pre-training strategy. It would be interesting to investigate whether CNN could achieve a comparable performance by developing an
	appropriate masked pre-training method.

	5. \textbf{Can the proposed method be utilized in novel clinical scenarios? } 
	In recent years, the field of ECG has experienced remarkable advancements, particularly in the area of deep learning. Numerous studies have revealed the ability of deep learning models to identify subtle changes in ECG signals that are unrecognizable by the human eye \citep{siontis2021artificial}. Remarkably, these subtle patterns may be indicative of asymptomatic diseases, such as asymptomatic left ventricular systolic dysfunction  \citep{attia2019screening, attia2022prospective, yao2021artificial}.  
	The detection of these patterns introduces novel clinical scenarios for ECG analysis \citep{somani2021deep}, 
	where the labels for ECG are obtained from alternative modalities such as echocardiogram and computed tomography. However, a significant challenge lies in the large number of ECG signals that lack corresponding labels from other modalities.  Previous research has typically excluded these unlabeled ECG records \citep{siontis2021artificial, somani2021deep}.  Our proposed method, characterized by its self-supervised nature, could effectively utilize these unlabeled ECG records. 
	Consequently, it becomes intriguing to explore the performance of the proposed method in such scenarios.
	

	\section{Conclusion}
	\label{sec:con}

	In conclusion, the paper presents a useful masked Transformer method which expands the application of MAE to ECG time series. We interpret an ECG time series as a sequence of segments and process it by the lightweight Transformer.
		Despite its simplicity, this method performs surprisingly well when adopting masked pre-training combined with proper training strategies. Therefore, the proposed algorithm outperforms recent state-of-the-art algorithms across multiple ECG classification datasets. In addition, the derived lightweight model offers deployment-friendly features, which is attractive in the clinical environment. 
		We hope that this study could direct future research on the application of Transformer to more ECG tasks.

\bibliographystyle{rss} 
\bibliography{reference}

\end{document}